\documentclass[useAMS,usenatbib]{mn2e}
\usepackage{xspace}
\usepackage{graphicx}
\newcommand{\ignore}[1]{}
\providecommand{\ao}{}
\renewcommand{\ao}{adaptive optics (AO)\renewcommand{\ao}{AO\xspace}\renewcommand{\Ao}{AO\xspace}\xspace}
\newcommand{\Ao}{Adaptive optics (AO)\renewcommand{\ao}{AO\xspace}\renewcommand{\Ao}{AO\xspace}\xspace}
\newcommand{\wfs}{wavefront sensor (WFS)\renewcommand{\wfs}{WFS\xspace}\renewcommand{\wfss}{WFSs\xspace}\xspace}
\newcommand{\wfss}{wavefront sensors (WFSs)\renewcommand{\wfs}{WFS\xspace}\renewcommand{\wfss}{WFSs\xspace}\xspace}
\newcommand{\shwfs}{Shack-Hartmann \wfs (SHWFS)\renewcommand{\shwfs}{SHWFS\xspace}\xspace}
\newcommand{\dm}{deformable mirror (DM)\renewcommand{\dm}{DM\xspace}\renewcommand{\dms}{DMs\xspace}\renewcommand{\Dms}{DMs\xspace}\renewcommand{\Dm}{DM\xspace}\xspace}
\newcommand{\dms}{deformable mirrors (DMs)\renewcommand{\dm}{DM\xspace}\renewcommand{\dms}{DMs\xspace}\renewcommand{\Dms}{DMs\xspace}\renewcommand{\Dm}{DM\xspace}\xspace}
\newcommand{\Dms}{Deformable mirrors (DMs)\renewcommand{\dm}{DM\xspace}\renewcommand{\dms}{DMs\xspace}\renewcommand{\Dms}{DMs\xspace}\renewcommand{\Dm}{DM\xspace}\xspace}
\newcommand{\Dm}{Deformable mirror (DM)\renewcommand{\dm}{DM\xspace}\renewcommand{\dms}{DMs\xspace}\renewcommand{\Dms}{DMs\xspace}\renewcommand{\Dm}{DM\xspace}\xspace}

\newcommand{\shs}{Shack-Hartmann sensor (SHS)\renewcommand{\shs}{SHS\xspace}\renewcommand{\shss}{SHSs\xspace}\xspace}
\newcommand{\shss}{Shack-Hartmann sensors (SHSs)\renewcommand{\shs}{SHS\xspace}\renewcommand{\shss}{SHSs\xspace}\xspace}
\newcommand{\lgs}{laser guide star (LGS)\renewcommand{\lgs}{LGS\xspace}\renewcommand{\lgss}{LGSs\xspace}\xspace}
\newcommand{\lgss}{laser guide stars (LGSs)\renewcommand{\lgs}{LGS\xspace}\renewcommand{\lgss}{LGSs\xspace}\xspace}
\newcommand{\ngs}{natural guide star (NGS)\renewcommand{\ngs}{NGS\xspace}\renewcommand{\ngss}{NGSs\xspace}\xspace}
\newcommand{\ngss}{natural guide stars (NGSs)\renewcommand{\ngs}{NGS\xspace}\renewcommand{\ngss}{NGSs\xspace}\xspace}
\newcommand{\mems}{Micro-Electro-Mechanical Systems (MEMS)\renewcommand{\mems}{MEMS\xspace}\xspace}
\newcommand{\snr}{signal to noise ratio (SNR)\renewcommand{\snr}{SNR\xspace}\xspace}
\newcommand{\moao}{multi-object \ao (MOAO)\renewcommand{\moao}{MOAO\xspace}\xspace}
\newcommand{\mcao}{multi-conjugate adaptive optics (MCAO)\renewcommand{\mcao}{MCAO\xspace}\xspace}
\newcommand{\ltao}{laser tomographic adaptive optics (LTAO)\renewcommand{\ltao}{LTAO\xspace}\xspace}
\newcommand{\cpu}{central processing unit (CPU)\renewcommand{\cpu}{CPU\xspace}\renewcommand{\cpus}{CPUs\xspace}\xspace}
\newcommand{\cpus}{central processing units (CPUs)\renewcommand{\cpu}{CPU\xspace}\renewcommand{\cpus}{CPUs\xspace}\xspace}
\newcommand{\psf}{point spread function (PSF)\renewcommand{\psf}{PSF\xspace}\renewcommand{\psfs}{PSFs\xspace}\xspace}
\newcommand{\psfs}{point spread functions (PSFs)\renewcommand{\psf}{PSF\xspace}\renewcommand{\psfs}{PSFs\xspace}\xspace}
\newcommand{\fpga}{field programmable gate array (FPGA)\renewcommand{\fpga}{FPGA\xspace}\renewcommand{\fpgas}{FPGAs\xspace}\xspace}
\newcommand{\fpgas}{field programmable gate arrays (FPGAs)\renewcommand{\fpga}{FPGA\xspace}\renewcommand{\fpgas}{FPGAs\xspace}\xspace}
\newcommand{\sor}{successive over-relaxation (SOR)\renewcommand{\sor}{SOR\xspace}\xspace}
\newcommand{\fdpcg}{Fourier domain pre-conditioned gradient (FDPCG)\renewcommand{\fdpcg}{FDPCG\xspace}\xspace}
\newcommand{\map}{maximum a-posteriori (MAP)\renewcommand{\map}{MAP\xspace}\xspace}
\newcommand{\elt}{Extremely Large Telescope (ELT)\renewcommand{\elt}{ELT\xspace}\renewcommand{\elts}{ELTs\xspace}\xspace}
\newcommand{\elts}{Extremely Large Telescopes (ELTs)\renewcommand{\elt}{ELT\xspace}\renewcommand{\elts}{ELTs\xspace}\xspace}
\newcommand{\dugall}{Durham University generalised adaptive optics laser laboratory (DUGALL)\renewcommand{\dugall}{DUGALL\xspace}\xspace}
\newcommand{\fwhm}{full-width at half-maximum (FWHM)\renewcommand{\fwhm}{FWHM\xspace}\xspace}
\newcommand{\wht}{William Herschel Telescope (WHT)\renewcommand{\wht}{WHT\xspace}\xspace}
\newcommand{\emccd}{electron multiplying CCD (EMCCD)\renewcommand{\emccd}{EMCCD\xspace}\xspace}
\newcommand{\dasp}{Durham \ao simulation platform (DASP)\renewcommand{\dasp}{DASP\xspace}\xspace}
\newcommand{\eelt}{European \elt (E-ELT)\renewcommand{\eelt}{E-ELT\xspace}\xspace}
\newcommand{\mpi}{Message Passing Interface (MPI)\renewcommand{\mpi}{MPI\xspace}\xspace}
\newcommand{\smp}{symmetric multi-processing (SMP)\renewcommand{\smp}{SMP\xspace}\xspace}
\newcommand{\svd}{singular value decomposition (SVD)\renewcommand{\svd}{SVD\xspace}\xspace}
\newcommand{\gpu}{graphical processing unit (GPU)\renewcommand{\gpu}{GPU\xspace}\renewcommand{\gpus}{GPUs\xspace}\xspace}
\newcommand{\gpus}{graphical processing units (GPUs)\renewcommand{\gpu}{GPU\xspace}\renewcommand{\gpus}{GPUs\xspace}\xspace}
\newcommand{\fft}{fast Fourier transform (FFT)\renewcommand{\fft}{FFT\xspace}\xspace}
\newcommand{\ifu}{integral field unit (IFU)\renewcommand{\ifu}{IFU\xspace}\xspace}
\newcommand{\darc}{the Durham adaptive optics real-time controller (DARC)\renewcommand{\darc}{DARC\xspace}\renewcommand{\Darc}{DARC\xspace}\xspace}
\newcommand{\Darc}{The Durham adaptive optics real-time controller (DARC)\renewcommand{\darc}{DARC\xspace}\renewcommand{\Darc}{DARC\xspace}\xspace}
\newcommand{\cots}{commercial off-the-shelf (COTS)\renewcommand{\cots}{COTS\xspace}\xspace}
\newcommand{\rtcp}{real-time control pipeline (RTCP)\renewcommand{\rtcp}{RTCP\xspace}\xspace}
\newcommand{\rms}{root-mean-square (RMS)\renewcommand{\rms}{RMS\xspace}\xspace}
\newcommand{\sFPDP}{serial Front Panel Data Port (sFPDP)\renewcommand{\sFPDP}{sFPDP\xspace}\xspace}
\newcommand{\wpu}{wavefront processing unit (WPU)\renewcommand{\wpu}{WPU\xspace}\xspace}
\newcommand{\canary}{CANARY\xspace}

\newcommand{\rtcs}{real-time control system (RTCS)\renewcommand{\rtcs}{RTCS\xspace}\xspace}
\newcommand{\ptp}{point-to-point (PTP)\renewcommand{\ptp}{PTP\xspace}\xspace}
\newcommand{\sse}{streaming SIMD extension (SSE)\renewcommand{\sse}{SSE\xspace}\xspace}
\newcommand{\api}{application programming interface (API)\renewcommand{\api}{API\xspace}\xspace}
\newcommand{\corba}{Common Object Request Broker Architecture (CORBA)\renewcommand{\corba}{CORBA\xspace}\xspace}
\newcommand{\lqg}{linear quadratic gaussian (LQG)\renewcommand{\lqg}{LQG\xspace}\xspace}
\newcommand{\scao}{single conjugate adaptive optics (SCAO)\renewcommand{\scao}{SCAO\xspace}\xspace}
\newcommand{\dma}{direct memory access (DMA)\renewcommand{\dma}{DMA\xspace}\xspace}
\newcommand{\xao}{extreme adaptive optics (XAO)\renewcommand{\xao}{XAO\xspace}\xspace}
\newcommand{\vlt}{Very Large Telescope (VLT)\renewcommand{\vlt}{VLT\xspace}\xspace}
\newcommand{\sparta}{Standard Platform for Advanced Real-Time
  Applications (SPARTA)\renewcommand{\sparta}{SPARTA\xspace}\xspace}
\newcommand{\eso}{European Southern Observatory (ESO)\renewcommand{\eso}{ESO\xspace}\xspace}
\newcommand{\eagle}{EAGLE\xspace}
\newcommand{\epics}{Exo-Planet Imaging Camera and Spectrograph (EPICS)\renewcommand{\epics}{EPICS\xspace}\xspace}
\newcommand{\iir}{infinite impulse response (IIR)\renewcommand{\iir}{IIR\xspace}\xspace}
\newcommand{\gtc}{Gran Telescopio Canarias (GTC)\renewcommand{\gtc}{GTC\xspace}\xspace}

\title[DARC: Capability and ELT suitability]{The Durham adaptive optics real-time controller:  Capability and ELT suitability}
\author[A. G. Basden, R. M. Myers]{A. G. Basden$^{1}$\thanks{E-mail: a.g.basden@durham.ac.uk (AGB)}, R. M. Myers$^1$\\
$^{1}$Department of Physics, South Road, Durham, DH1 3LE, UK}

\begin{document}
\maketitle
%\title{The Durham adaptive optics real-time controller:  Capability  and ELT suitability}

%\tableofcontents
%\clearpage
%\title{The Durham adaptive optics real-time controller:  Capability and ELT suitability}

%\author[A. G. Basden and
%  R. M. Myers]{A. G. Basden$^{1}$\thanks{E-mail:
%    a.g.basden@durham.ac.uk (AGB)} and R. M. Myers$^1$\\
%$^{1}$Department of Physics, South Road, Durham, DH1 3LE, UK}
%\author{A. G. Basden and R. M. Myers}
%\address{Department of Physics, South Road, Durham, DH1 3LE, UK}
%\email{a.g.basden@durham.ac.uk}

%\begin{document}
%\maketitle
%\bsp

\begin{abstract}
The Durham adaptive optics real-time controller is a generic, high
performance real-time control system for astronomical adaptive optics
systems.  It has recently had new features added as well as
performance improvements, and here we give details of these, as well
as ways in which optimisations can be made for specific adaptive
optics systems and hardware implementations.  We also present new
measurements that show how this real-time control system could be used
with any existing adaptive optics system, and also show that when used
with modern hardware, it has high enough performance to be used with
most Extremely Large Telescope adaptive optics systems.
\end{abstract}
%\ocis{(010.1080)   Active or adaptive optics, (110.1080)   Active or
%  adaptive optics, (200.4960)   Parallel processing }

%\bibliographystyle{osajnl.bst}
%\bibliography{mybib}

\section{Introduction}
\Ao \citep{adaptiveoptics} is a technique for mitigating the degrading
effects of atmospheric turbulence on the image quality of ground-based
optical and near-IR telescopes. It is critical to the high angular
resolution performance of the next generation of \elt facilities,
which will have primary mirror diameters of up to 40~m. Without
mitigation, the general spatial resolution of such a telescope would
be subject to the same atmospheric limitations as a 0.5~m diameter
telescope. The proposed \elts represent a large strategic investment
and their successful operation depends on having a range of high
performance heterogeneous \ao systems.  As such, these telescopes will
be the premium ground based optical and near-IR astronomical
facilities for the next two decades.  The \elts will, however, require
a very significant extrapolation of the \ao technologies currently
deployed or under development for existing 4--10~m telescopes.  Not
least amongst the required developments of current \ao technology is
the area of real-time control. In this paper we describe the
development and testing of a real-time controller, with the required
scalability.

\Darc \citep{basden9} is a real-time control system for \ao which was
initially developed to be used with the \canary on-sky \moao
technology demonstrator \citep{canaryresultsshort}.  As such, it was a
significant success, being stable, configurable and powerful, and able
to meet all the needs for this \ao system.  There was also demand for
\darc to be used with other instruments, and so a further improved
version of \darc was released to the public using an open source
GNU-GPL \citep{gpl} license.  Here, we provide information about the
\darc platform, including new features, architectural changes,
modularisation, performance estimates, algorithm implementation and
generalisations.  We have tested the performance of this system in
configurations matching a wide range of proposed \elt \ao systems, and
also configured to match proposed high order 8~m class telescope \ao
systems, and a selection of results are presented here.  

To date, each \ao system commissioned on a telescope or used in a
laboratory has generally had its own real-time control system, leading
to much duplicated effort.  So far as we are aware the only other {\em
  multi-use} high performance real-time control system for \ao is the
\eso \sparta \citep{sparta}. Like \darc, \sparta is designed to
support heterogeneous components in high performance configurations,
including computational hardware other than standard PCs, and is
currently being integrated with second generation \vlt instruments
such as SPHERE \citep{sphere}, GRAAL \citep{2010SPIE.7736E..57P} and
GALACSI \citep{2006NewAR..49..618S}, though is not used outside \eso.
The full \sparta system cannot be used with just standard PCs and so
expert programming skills are required, as well as dedicated software
maintenance, and so would be unsuitable and costly for simple
laboratory setups.  A solution to this is to develop a system that is
flexible enough to satisfy most performance requirements, is able to
meet challenging \ao system specifications using standard PC hardware,
is simple to set up and use, and also supports hardware acceleration
so that it is powerful enough to be used with demanding applications
on-sky.  \darc has been designed with these requirements in mind, and
here we seek to demonstrate how it can be suited for most \ao systems.
A common \ao real-time control system of this kind would be beneficial
for the \ao community, leading to a reduction in learning time, and
increased system familiarisation.

This \cpu-based approach to \ao real-time control has not been
successful in the past because it has been deemed that previously
available \cpus have not been sufficiently powerful to meet the
demanding performance requirements of on-sky \ao systems
\citep{sparta}.  The advent of multi-core processors, which started to
become commonly available from the mid-2000s, has been a key enabling
factor for allowing our approach to succeed (our development commenced
in 2008).  It is now possible to obtain standard PC hardware with
enough \cpu cores to not only perform the essential real-time pipeline
calculations (from wavefront sensor data to \dm commands), but also to
perform necessary sub-tasks, such as parameter control, configuration,
and sharing of real-time system information and diagnostic streams.
Attempts at using a single core \cpu for all these tasks have
generally failed because context switching between these tasks has led
to unacceptable jitter in the \ao system.  However, multi-core systems
do not suffer so much from this unpredictability.  Previous systems
have also not been freely available and have been closed source, which
has greatly hindered uptake particularly for laboratory bench based
systems.  Our approach does not have these restrictions.

Commercially available offerings, although impressive in many
respects, do not scale well for use with future high order \ao system
designs, and are often restricted to specific hardware, are designed
for laboratory systems and can lack features required for high order
on-sky \ao systems, such as pipe-lined pixel-stream processing.  

There are three main areas of application for \darc.  As a laboratory
\ao \rtcs where flexibility and modularity are key, as well as
stability.  As a control system for instruments on 8--10~m class
telescopes, where it is being evaluated for use with a number of \ao
systems currently under development.  Finally, as a control system for
\elt instruments, where it is a currently a potential candidate for
use with two proposed \ao systems.

\ignore{both MAORY \citep{2010SPIE.7736E..99F} and EAGLE \citep{eagleScience}}

In \S2, we give an overview of the new features, including advanced
algorithms, modularisation, diagnostic data handling and tools
available with \darc.  In \S3, we discuss how \darc can be optimised
for use with a given \ao system, and present some results
demonstrating this optimisation.  In \S4, we provide some examples of
how \darc can be used with some existing and proposed demanding \ao
systems, and demonstrate the hardware that would be required for such
operation.  Finally, in \S5 we draw our conclusions.

\section{DARC features and algorithms}

There are several important architectural changes that have been made
between the original version of \darc \citep{basden9} as used on-sky
with \canary and the current freely available version (to be used with
future phases of \canary) and these are are discussed here.  The
changes include improved and expanded modularisation, changes to
diagnostic data handling, improvements to \gpu acceleration, the
ability to be used asynchronously with multi-rate cameras, a
generalisation of pixel handling, advanced spot tracking algorithms,
improved command line tools and the ability to use \darc in a
massively parallelised fashion across multiple nodes in a computing
grid.  Hardware acceleration support has been improved principally
through the increased modularisation of \darc, and overall performance
has also been dramatically improved.

\darc has also acquired the ability to allow user parameter changes on
a frame-by-frame basis, allowing more complete control and dynamic
optimisation of the \ao system.  The control interface has added
functionality that includes parameter subscription and notification,
and greater control of diagnostic data, including redirection, partial
subscription and averaging.

\subsection{DARC modularisation}
Since conception, \darc has always had some degree of modularisation;
it has been possible to change cameras, deformable mirrors and
reconstruction algorithms be dynamically loading and unloading modules
into the real-time pipeline, without restarting \darc.  This
modularisation has been extended to increase the degree of user
customisation that is possible with \darc.  Module interfaces that
allow modules to be dynamically loaded and unloaded in \darc now
additionally include image calibration and wavefront slope computation
interfaces, and an asynchronous open-loop \dm figure sensing interface, as
well as the pre-existing wavefront reconstruction interface.  A
parameter buffer interface has also been added, allowing customisation
of high-speed parameter input, facilitating the adjustment of any
parameter on a frame-by-frame basis, thus allowing advanced use of the
system, for example \dm modulation and fast reference slope updating.  

The user \api for developing \darc modules has been rationalised and
now most modules include similar functions, which are called at well
defined points during the data-processing pipeline.  This allows the
developer to easily identify which functions are necessary for
them to implement to achieve optimum \ao loop performance, and
encourages the consideration of algorithms that reduce latency, and
improve real-time performance.

Although it is possible to implement a large number of functions in
each of these modules, typically they are not all required and so the
developer should implement only the necessary subset for their
particular application. 

\subsubsection{Module hook points}
\begin{figure}
\includegraphics[width=8cm]{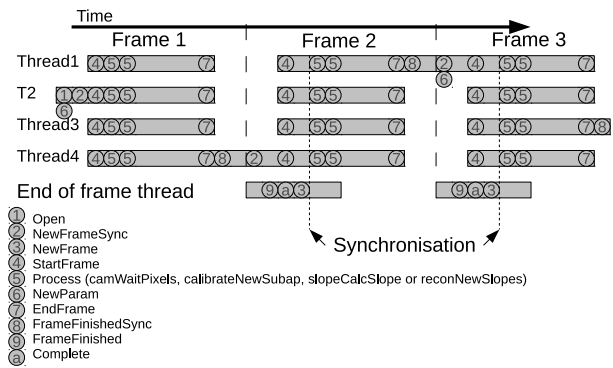}
\caption{A figure demonstrating the points (in time) at which DARC
  module functions, if implemented, are called.  Three WFS frames are
  shown with time increasing from left to right, with a new module
  being loaded at the start (point 1).  This assumes there are four
  main processing threads (Thread1, T2, Thread3 and Thread4), and the
  thread responsible for pre-processing and post-processing is also
  shown (``End of frame thread'').  The legend shows the function
  names corresponding to the points (in time) at which the functions
  are called, for example, (1) is the point at which the module
  ``Open'' function is called, thus initialising the module.  The
  process function (5) is called repeatedly each frame until the frame
  has been processed (for example, once for each sub-aperture).
  Further explanation is given in the main text.}
\label{fig:moduleHooks}
\end{figure}

\Darc uses a horizontal processing strategy as described by
\citet{basden9}, which splits computational load as evenly as possible
between available threads, thus allowing good \cpu load balancing and
high \cpu resource utilisation, giving low latency performance.  To
achieve this, each thread must be responsible for performing multiple
algorithms, including \wfs calibration, slope computation and partial
wavefront reconstruction (rather than separate threads performing
calibration, slope computation and reconstruction).  To reconcile this
with the modular nature of \darc, there are defined points at which
module functions are called as shown in Fig.~\ref{fig:moduleHooks}.  A
module developer then fills in the body of the module functions that
they require.  \darc will then call these functions at the appropriate
time in a thread-safe way.  Fig.~\ref{fig:moduleHooks} shows the \darc
threading structure and the points at which module functions are
called.  It should be noted that the ``Process'' function is called
multiple times for each \wfs frame until all the data have been
processed (i.e. called for each sub-aperture in a Shack-Hartmann
system).  This approach has been taken to encourage a module developer
to consider how their algorithm best fits into a low latency
architecture, and to provide a consistent interface between modules.
Unimplemented functions (those that are not required for a given
algorithm) are simply ignored.

We make a distinction between processing threads (labelled Thread1,
Thread3 etc.\ in Fig.~\ref{fig:moduleHooks}) which do the majority of
the parallelised workload, and the ``end-of-frame'' thread which is
used to perform sequential workloads such as sending commands to a
deformable mirror.  Upon initialisation of a module, the module
``Open'' function is called by a single processing thread.  Here, any
initialisation required is performed, such as allocating necessary
memory, and (for a camera module) initialising cameras.  Parameters
(such as a control matrix, or camera exposure time) are passed to the
module using the \darc parameter buffer.  To access this buffer, the
``NewParam'' function is then called (if implemented).  After this,
the module is ready to use.  One processing thread calls a
``NewFrameSync'' function, for per-frame initialisation.  Each
processing thread then calls a ``StartFrame'' function, for
per-thread, per-frame initialisation.  The ``Process'' function is
then called multiple times, while there is still data to be processed
(for a Shack-Hartmann system this is typically once per sub-aperture,
shared between the available processing threads).  Once all such
processing has finished, each processing thread calls an ``EndFrame''
function, which would typically perform gather operations to collate
the results (for example summing together partial \dm vectors).  A
single thread is then chosen to call a ``FrameFinishedSync'' function
to finalise this frame.  After this function has been called, the
``end-of-frame'' thread springs into life, allowing the processing threads
to begin processing of the next frame, starting with the
``NewFrameSync'' function.  The ``end-of-frame'' thread calls a
``FrameFinished'' function for finalisation during which (for a mirror
module) commands should be send to the \dm.  A ``Complete'' function
is then called for each module, which can be used for initialisation
ready for the next frame.  The ``end-of-frame'' thread then calls a
``NewFrame'' function.  The ``end-of-frame'' thread is not
synchronised with the processing threads, and so there is no guarantee
when the ``NewFrame'' function is called relative to the functions
called by the processing threads, except that processing threads will
block before calling the ``Process'' function until the
``end-of-frame'' thread has finished.

Whenever the \darc parameter buffer is updated, the ``NewParam''
function will be called by a single processing thread just before the
``NewFrameSync'' function is called.  When the module is no longer
required (for example when the user wishes to try a new algorithm
available in another module), a ``Close'' function is called, to free
resources.

The large number of functions may seem confusing, particularly since
some appear to have similar functionality.  Fortunately, most modules
need only implement a small sub-set of these functions.  The full
suite of functions have been made available to give a module developer
the required flexibility to create a module that is as efficient as
possible, minimising \ao system latency.

Matrix operations are highly suited to this sort of horizontal
processing strategy since they can usually be highly parallelised, and
thus divided between the horizontal processing threads.  Wavefront
reconstruction using a standard matrix-vector multiplication algorithm
(with a control matrix) is therefore ideally suited.  Iterative
wavefront reconstruction algorithms, for example those based around
the conjugate gradient algorithms are less easy to parallelise, since
each iteration step depends on the previous step.  However, a
horizontal processing strategy does allow the first iteration to be
highly parallelised, which can lead to significant performance
improvements when the number of iterations is small, for example when
using appropriate preconditioners such as those used in the fractal
iterative method \citep{2006SPIE.6272E..90B}.  The post-processing
(end of frame) thread (or threads) can then be used to compute the
remaining iterations.  Similarly, any system that requires multiple
step reconstruction, for example a projection between two vector
spaces, such as for true modal control, can be easily integrated.

\subsubsection{DARC camera modules}
An example of several simple camera modules are provided with the
\darc source code.  These are modules for which camera data are only
available on a frame-by-frame basis (rather than a pixel-by-pixel
basis), and typically would be used in a laboratory rather than
on-sky.  In this case, the camera data are transferred to \darc at the
start of each frame (using the ``NewFrameSync'' function), and other
functions are not implemented (except for ``Open'' and ``Close'').
For such cameras, there is no interleaving of camera readout and pixel
processing, and so a higher latency results.

For camera drivers which have the ability to provide pixel stream
access (i.e.\ the ability to transfer part of a frame to the computer
before the detector readout has finished), a more advanced camera
module can be implemented, and examples are provided with \darc.
Such modules allow interleaving of camera readout with processing,
making use of the ``Process'' function to block until enough pixels
have arrived for a given \wfs sub-aperture, after which calibration and
computation of this sub-aperture can proceed.  

\subsubsection{Parameter updating}
\darc has the ability to update any parameter on a frame-by-frame
basis.  However, since this could mean a large computational
requirement (to compute the parameters from available data), or a
large data bandwidth requirement (for example for updating a control
matrix), this update ability is implemented using a \darc module
interface.  The user can then create such a module depending on their
specifications to best suit the needs and requirements of their system,
for example using a proprietary interconnect to send the parameters.
A standard set of \darc functions are provided, which should
be overwritten for this parameter buffer interface to take effect.
This buffer module is again dynamically loadable, meaning that this
ability can easily be switched on and off.

This ability to update parameters every frame or in a deterministic
fashion (at a given frame number for example) is optional, and
additional to the non-real-time parameter update facility that is
used to control \darc as discussed by \citet{basden9}, which allows
parameters to be changed in a non-deterministic fashion (non-real-time
i.e.\ there is no guarantee that a parameter will be changed at a
particular frame number or time).

\subsection{Diagnostic data handling}
The general concept of diagnostic data handling is described in the
original \darc paper \citep{basden9}: In summary, there is a separate
diagnostic stream for each diagnostic data type (raw pixel data,
calibrated pixel data, wavefront slope measurements, mirror demands,
etcetera).  It should be noted that diagnostic data handling is used
only to provide data streams to clients, not for the transfer of data
along the real-time pipeline.  As such, the diagnostic data system
does not need to be hard-real-time.  

Transport of \darc diagnostic data uses TCP/IP sockets by default,
which lead to a reliable, simple and fairly high performance system
that is easy to understand and set up, with minimal software
configuration and installation.  However, because \darc is modular by
design, users are able to replace this system with their own should
they have a need to, using their own transport system and protocols to
distribute this data to clients.  This is well suited to a facility
class telescope environment, where standardised protocols must be
followed.

\subsubsection{Default diagnostic data implementation}
\begin{figure}
\includegraphics[width=8cm]{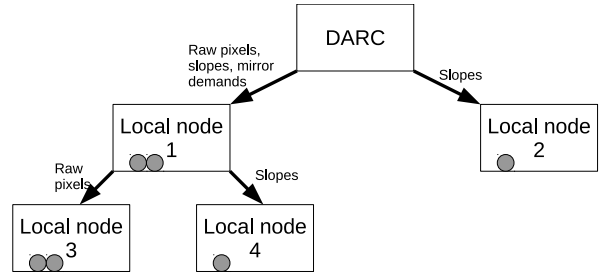}
\caption{A figure demonstrating the DARC diagnostic data system.
Here, raw pixel, slope and mirror data are being sent from the
real-time part of the system to one local node, which is in turn
sending raw pixels and slopes to other nodes.  A further local node is
receiving slopes directly from the real-time part of the system.
Clients which process these data are represented by grey circles.}
\label{fig:telemetry}
\end{figure}

The default \darc diagnostic system seeks to minimise network
bandwidth as much as possible using \ptp connections.  Diagnostic data
are sent from the real-time system to a remote computer, where the
data are written into a shared memory circular buffer.  Clients on
this computer can then access the data by reading from the circular
buffer, rather than requesting data directly from the real-time system
across the network.  Additionally, these data can then be
re-distributed to further remote computers, allowing the data to be
read by other clients here, as shown in Fig.~\ref{fig:telemetry}.
Hence, each diagnostic stream needs to be sent only once (or,
depending on network topology, a small number of times) from the main
real-time computer (rather than once per client), and network
bandwidth can be tightly controlled.  

In cases where only part of a diagnostic stream is required (for
example, pixel data from a single camera in a multi-camera system, or
a sub-region of an image), these data can be extracted by a \darc
client into a new diagnostic stream before being distributed over the
network, reducing bandwidth requirements.  Additionally, for cases
where only the sum or average of many frames of data from a given
stream is required this operation can be performed using a client
provided by \darc before data are transported over the network, again
greatly saving network bandwidth, for example for image calibration.

It should be noted that the default diagnostic system does not use
broadcasting or multicasting.  This is because broadcasting and (in
its simplest form) multicasting are inherently unreliable and thus
would not provide a reliable diagnostic stream, giving no guarantee
that data would reach their destination, which is undesirable for an
\ao system.  Although reliable multicasting protocols are available, a
decision has been made not to use these by default, because this would
increase the complexity of \darc, and would be unnecessary for users
of simple systems.  However we would like to reiterate that such
diagnostic systems can easily be added and integrated with \darc by
the end-user.

On the real-time computer and remote nodes with diagnostic clients, a
region of shared memory (in /dev/shm) is used for each diagnostic
stream, implemented using a self describing circular buffer typically
hundreds of entries long (though this is configurable and in practise
will depend on available memory, and stream size).  Streams can be
individually turned on and off as required using the \darc control
interface (using a graphical, script or command line client or the
\api interface).  \darc will write diagnostic data to the circular
buffers of streams that are not switched off, and these data can then
be read by clients or transported.  The rate at which \darc writes
these data can be changed (every frame, or every $n$ frames with a
different rate for each diagnostic stream).  Clients can then retrieve
as much or as little of these data as they require, by setting the
sub-sampling level at which they wish to receive.

Since TCP/IP is used, retransmission of data may be necessary when
network hic-cups occur (though this is handled by the operating
system).  The processes responsible for sending data over the network
to clients will block until an acknowledgement from the client has
been received (this is handled by the operating system), and therefore
at times may block for a longer than average period while waiting for
retransmission.  If this happens frequently enough (for example on a
congested network), then the head of the circular buffer (the location
at which new diagnostic data are written to the circular buffer) will
catch up with the tail of the buffer (the location at which data are
sent from).  To avoid data corruption, the \darc sending process will
jump back to the head of the circular buffer once the tail of the
buffer falls more than 75~\% of the buffer behind the head.
Therefore, a chunk of frames will be lost.  This is undesirable, but
should be compared with what would occur with an unreliable protocol,
for example UDP.  Here, the sending process would not be blocked, and
so would keep up with the head of the circular buffer.  However when a
packet fails to reach its destination, due to congestion or a network
hic-cup, the packet would simply be dropped and not retransmitted.
Therefore, there would be a portion of a frame of data missing
(corresponding to the dropped packet), rendering (in most cases) the
entire frame unusable.  We should therefore consider two cases: In a
highly loaded network, it is likely that the number of partial
(unusable) frames received would be greater than the number of frames
dropped when using a reliable protocol even though the total network
throughput would be greater, because dropped packets would be
dispersed more-or-less randomly affecting a greater number of frames,
while dropped frames would be in chunks.  In a less loaded network,
UDP packets would still be occasionally dropped, resulting in unusable
frames, while a reliable protocol would (on average) be able to keep
up with the circular buffer head, dropping behind occasionally, but
not far enough to warrant dropping a chunk of frames to jump back to
the circular buffer head.

\subsubsection{Disadvantages}
The disadvantage of this simple approach to diagnostic data is that
TCP/IP is unicast and \ptp, meaning that if multiple clients on
different computers are interested in the same data then the data are
sent multiple times, a large overhead.  However, we take the view that
the simplicity of the default system out-weights the disadvantages,
and that for more advanced systems, a separate telemetry system should
be implemented, for which, we are unable to anticipate the requirements.

\subsubsection{Diagnostic feedback}
We have so far discussed the ability to propagate real-time data to
interested clients.  However, it is often the case that these data will
be processed and then injected back into the real-time system on a
per-frame basis, useful not only for testing, but also for calibration
tools such as turbulence profiling, or deformable mirror shape
feedback.  This feedback interface module can be implemented in \darc in
several ways.  One option is to use the per-frame parameter update
module interface as discussed previously, provided by the user to suit
their requirements.  Another option is to modify an existing \darc
processing module (dynamically loadable), to accept the expected
input, for example modify a wavefront reconstruction module to accept
an additional input (via Infiniband or any desired communication
protocol) of actual \dm shape, to be used for pseudo-open-loop control
reconstruction.

\subsection{User facilities}
The \darc package comes with an extensive suite of user tools,
designed to simplify the setup and configuration of \darc.  These
include command-line based tools, a graphical interface, a
configurable live display tool and an \api.  Since the \darc control
interface is based upon \corba \citep{corba}, a client can be written in any
programming language which has a suitable \corba implementation.

Using these tools allows additional customised packages and facilities
to be built, specific for the \ao system in question.  An example of
this would be a tip-tilt offload system, which would capture slope
diagnostic data from \darc, and if mean slope measurements became too
large would inform the telescope to update its tracking.  Many such
systems based around \darc diagnostic data have been used successfully
with \canary.

\subsection{Configurable displays}
\darc does not know the nature of the data in the diagnostic streams
that it produces.  It is known, for example, that a particular stream
contains \dm demand data, and how many \dm actuators there are;
however, \darc knows nothing about the mapping of these actuators onto
physical \dms, how many \dms there are, and what geometry they have.
Therefore, to display this data in a way meaningful to the user,
additional information is required.  

The \darc live display tool allows user configuration, both manually
entered and via configuration files.  Additionally, a collection of
configuration files can be used and the live display will present
a selection dialogue box for the user to rapidly switch between
configurations, for example to display a phase map of different \dms,
or to switch between a phase map and a \wfs image display.
Each configuration can specify which diagnostic streams should be
received and at what rate, allowing a given configuration to receive
multiple diagnostic streams simultaneously, for example, allowing a
spot centroid position to be overlain on a calibrated pixel display.
Manually entered configuration is useful while an \ao system is being
designed and built, and can be used for example to change a one
dimensional pixel stream into a two dimensional image for display.  

The live display can be configured with user selectable buttons,
which can then be used to control this display configuration.  For
example, when configured for a \wfs pixel display, the user might be
able to turn on and off a vector field of the slope measurements by
toggling a button.  

\begin{figure}
\includegraphics[width=8cm]{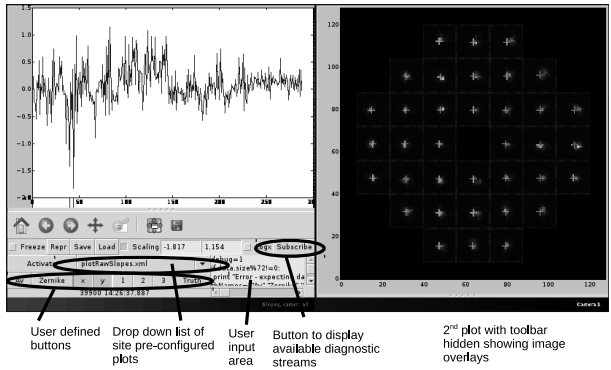}
\caption{A figure showing two instances of the DARC live display tool.
  On the left hand side a one dimensional display of slope
  measurements is shown, along with the associated toolbar, configured
  for this AO system (CANARY) allowing the user to select which slopes
  from which wavefront sensor to display.  On the right hand side, a
  two dimension wavefront sensor image is displayed, with sub-aperture
  boundaries and current spot locations overlain (as a grid pattern,
  and cross hairs respectively), and the display is receiving multiple
  diagnostic streams (image and slopes) simultaneously.  In this
  display, the toolbar is hidden (revealed with a mouse click).}
\label{fig:darcplot}
\end{figure}

This configurability is aimed at ease of use, as a simple way of
getting a user friendly \ao system up and running.  Although not
designed to be used as a facility class display tool, it does have
sufficient flexibility and capability to function as such.
Fig.~\ref{fig:darcplot} shows two instances of the display tool being
used to display \wfs slope measurements and calibrated pixel data.

\subsection{DARC algorithms}
Since \darc is primarily based around \cpu computation and is modular,
it is easy to implement and test new algorithms.  As a result, there
are a large number of algorithms implemented, and this list is
growing as new ideas or requirements arise.  Some of these algorithms
are given by \citet{basden9}, and we present additional algorithms here.

\subsubsection{Pixel processing and slope computation}
The use of a brightest pixel selection algorithm \citep{basden10} has
been demonstrated successfully on-sky using \darc.  This algorithm
involves selecting a user determined number of brightest pixels in
each sub-aperture and setting the image threshold for this
sub-aperture at this level.  This helps to reduce the impact of
detector readout noise, and can lead to a significant reduction in
wavefront error.

The ability of \darc to perform adaptive windowing, or spot tracking
with Shack-Hartmann wavefront sensors, has been mentioned previously
\citep{basden9}.  Further functionality has now been added allowing
groups of sub-apertures to be specified, for which the adaptive window
positions are computed based on the mean slope measurements for this
group, and hence these grouped sub-apertures all move together.  This
allows, for example, tip-tilt tracking on a per-camera basis when
multiple wavefront sensors are combined onto the same detector.  

One danger with adaptive windowing is that in the event of a spurious
signal (for example a cosmic ray event) or if the signal gets lost
(for example intermittent cloud), then the adaptive windows can move
away from the true spot location.  Adaptive window locations will then
be updated based upon noise, and so the windows will move randomly
until they find their Shack-Hartmann spot, or fix on another nearby
spot.  To prevent this from happening, it is possible to specify the
maximum amount by which each window is allowed to move from the
nominal spot location.  Additionally, adaptive window locations are
computed from the local spot position using an \iir filter, for which
the gain can be specified by the user, helping to reduce the
impact of spurious signals.

In addition to weighted centre of gravity and correlation based slope
computation, a matched filter algorithm can also be used.  \darc can
be used not only with Shack-Hartmann wavefront sensors, but also with
Pyramid wavefront sensors and, in theory, with curvature wavefront
sensors (though this has never been tested), due to the flexible
method by which pixels are assigned to sub-apertures (which can be
done in an arbitrary fashion).  The modular nature of \darc means that
other sensor types could easily be added, for example YAW \citep{yaw}
and optically binned Shack-Hartmann sensors \citep{basden7}.

\subsubsection{Reconstruction and mirror control}
In addition to matrix-vector based wavefront reconstruction (allowing
for least-squares and minimum mean square error algorithms), \lqg
reconstruction can also be carried out, allowing for vibration
suppression, which can lead to a significant performance improvement
\citep{lqg}.  An iterative solver based on preconditioned conjugate
gradient is also available, and can be used with both sparse and dense
systems.  Using this reconstruction technique \citep{pcg} has the
advantage that a matrix inversion to compute a control matrix is not
required.  In addition, an open-loop control formulation is available
which allows \dm commands ($a$) to be computed according to
\begin{equation}
a_i = (1-g) E\cdot a_{i-1} + g R\cdot s_i
\end{equation}
where $s_i$ are the current wavefront slope measurements, $g$ is a
gain parameter, $R$ is the control matrix, and $E$ is a square matrix,
which for an integrator control law would be equal to an identity
matrix scaled by $\frac{1}{1-g}$.  

\darc also contains the ability to perform automatic loop opening in
the case of actuator saturation, i.e. to automatically open the
control loop and flatten the \dm if a predefined number of actuators
reach a predefined saturation value.  This can be important to avoid
damage while testing new algorithms and control laws.

Some \dms display hysteresis and other non-linear behaviour that
results in the shaped \dm not forming quite the shape that was
requested.  To help reduce this effect, \darc includes the option to
perform actuator oscillation around the desired \dm shape position,
allowing the effect of hysteresis to be greatly reduced.  Typically, a
decaying sine wave is used, with the decay leading to the desired position.

Since \darc has the ability to update parameters on a frame-by-frame
basis, it has the ability to modulate (or apply any time-varying
signal to) some or all of the \dm actuator demands, allowing complex
system operations to be performed.

\subsubsection{Advanced computation}
\darc has the ability to operate multiple \wfss asynchronously,
i.e.\ at independent frame rates (which are not required to be
multiples of each other).  This gives the ability to optimise
wavefront sensor frame rate depending on guide star brightness, and so
can lead to an improvement of \ao system performance.  This capability
is achieved by using multiple instances of \darc, one for each \wfs,
which compute partial \dm commands based on the \wfs data, and a
further instance of \darc which combines the partial \dm commands once
they are ready, using shared memory and mutual exclusion locks for
inter-process communication.  The way in which the partial \dm
commands are combined is flexible, with the most common option being
to combine these partial commands together as they become available
and then update the \dm, meaning that the \dm is always updated with
minimal latency.

The ability to operate multiple instances of \darc, combined with
modularity, means that \darc can be used in a distributed fashion,
allowing computational load to be spread over a computing grid.  To
achieve this, modules responsible for distributing or collating data
at different points in the computation pipeline are used, with data
being communicated over the most efficient transport mechanism.  At
present, such modules exist based on standard Internet sockets, and
also using shared memory.  This flexibility allows \darc to be
optimised for available hardware.  Extremely demanding system
requirements can therefore be met.

\subsection{Hardware acceleration}
The modular nature of \darc makes it ideal for use with acceleration
hardware, such as \fpgas and \gpus.  Two hardware acceleration modules
currently exist for \darc (and of course more can easily be added).
These are an \fpga based pixel processing unit that can perform image
calibration and optionally wavefront slope calculation and is
described elsewhere \citep{sparta}, and a \gpu based wavefront
reconstruction module, which we now describe.

\subsubsection{GPU wavefront reconstruction}
The \gpu wavefront reconstruction module can be used with any CUDA
compatible \gpu.  It performs a matrix-vector
multiplication based wavefront reconstruction, which (depending on the
matrix) includes least squares and minimum variance reconstruction.
As discussed previously, \darc uses a horizontal processing strategy
processing sub-apertures as the corresponding pixel data become
available.  The \gpu reconstruction module also follows this strategy,
with partial reconstructions (using appropriate subsets of the matrix)
being performed before combination to provide the reconstructed
wavefront.  

The control matrix is stored in \gpu memory, and wavefront slope
measurements are uploaded to the \gpu every frame.  The \gpu kernel
that performs the operations has been written specifically for \darc,
providing a 70~\% performance improvement over the standard cuBlas
library which is available from the \gpu manufacturer NVIDIA. This
module uses single precision floating point data.

The performance reached by this module is limited by \gpu internal
memory bandwidth rather than computation power, since the matrix has
to be read from \gpu memory into \gpu processors every frame.  We are
able to reach about 70~\% of peak theoretical performance on a NVIDIA
Tesla 2070 \gpu card.  An alternative version of this \darc module
also exists which stores the control matrix in a 16 bit integer
format, with conversion to single precision floating point performed
each frame before multiplication.  This allows a performance
improvement of about 80~\% (reducing computation time by 45~\%) with a
trade-off of reduced precision.  However, as shown by \citet{basden8},
16 bit precision is certainly sufficient for some \elt scale \ao
systems (it may not be sufficient for a high contrast system, though we
have not investigated this).  It should be noted that slope
measurements for astronomical \ao systems are typically accurate to at
most 10--11 bits of precision, limited by photon noise.

\section{DARC optimisation}
\darc has been designed to provide low latency control for \ao,
operating with a baseline of minimal hardware (a computer), whilst
also providing the ability to use high end hardware, and hardware
acceleration.  This has been achieved by careful management of the
workload given to processor threads using a horizontal processing
strategy, and by reducing the need for synchronisation, as described
by \citet{basden9}.  \darc has since been updated to further reduce
this latency, with steps being taken to reduce the number of thread
synchronisation points (thus reducing synchronisation delays), and
also providing control over where post-processing is performed.  The
increased modularisation of \darc has led to a clearer code structure
and allowed the thread synchronisation points to be rationalised,
providing opportunities for better optimisation of modules, which can
be fitted into the \darc structure more easily.

\subsection{Site optimisation}
To optimise \darc for a specific application, there are several steps
that can be taken to allow best performance (lowest latency) to be
achieved for a given hardware setup.  For all of these steps, no
recompilation is necessary, and many can be performed without
stopping \darc.  In this section we will present these optimisations
and discuss why they can make a difference, and how they should be
applied.  The large number of optimisation points built into \darc
mean that it is well suited to meet the demands of future instruments.

\subsubsection{Number of threads}
\begin{figure}
\includegraphics[width=8cm]{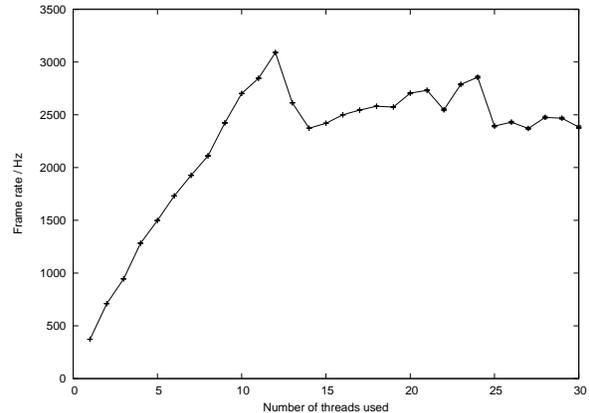}
\caption{A plot showing how AO system maximum achievable frame rate
  depends on the number of DARC processing threads used.
  Uncertainties are shown but generally too small to discern.}
\label{fig:nthreads}
\end{figure}

The main way to optimise \darc performance is to adjust the number of
processing threads used.  A higher order \ao system will have reduced
latency when more processing threads are used; however, the number of
threads should be less than the number of processing cores available,
which will in turn depend on the processing hardware.  The balance
between processing power and memory bandwidth can also affect the
optimal number of processing threads.  Fig.~\ref{fig:nthreads} shows
how maximum achievable frame rate is affected by the number of
processing threads for a $40\times40$ sub-aperture \scao system, and
it should be noted that a higher frame rate corresponds to lower
latency.  The frame rates displayed here were measured using a
computer with two 6-core Intel E5645 processors with a clock speed of
2.4~GHz and hyper-threading enabled, giving a total of 24 processing
cores (12 physical).  We have restricted threads to run on a single
core (i.e.\ no thread migration, by setting the thread affinity), with
the first six threads running on the first \cpu, the next six the
second \cpu, the next six on the first and so on as required (so each
processor core may have multiple threads, but each thread is only
allowed on a single core).  It can be seen from this figure that \darc
gives near linear performance scaling with number of threads up to the
number equal to the number of physical \cpu cores (12), at which
point, maximum performance is achieved.  This is followed by a dip at
13 and 14 threads as one core is then having to run two \darc threads.
Performance then increases again up to 24 threads (except for a dip at
22 threads, which we are unsure about, but which is repeatable), after
which performance levels off (and eventually falls) as each core has
to run an increasing number of threads, and thread synchronisation
then begins to take its toll.

This figure shows that hyper-threading is detrimental to performance
in this case, and we recommend that the use of hyper-threading should
be investigated by any \darc implementer and switched off (in the
computer BIOS) if it is detrimental.  However, this is not a condition
that we would wish to impose because all situations are different and
some users may find it desirable to have hyper-threading.

It should be noted that there is also the option of using a separate
thread for initialisation and post-processing of data, or to have this
work done by one of the main processing threads, and each option can
provide better performance for different situations.  

\subsubsection{Affinity and priority of threads}
By giving processing threads elevated priorities the Linux kernel will
put more effort into running these threads.  However, when using
\darc, best performance is not necessarily achieved by giving all
threads high (or maximum) priority.  Rather, there can be an advantage
in considering the work that each individual thread is required to do,
and whether particular threads are likely to be on the critical path
at any given point.

As a simple example, consider the case of an \ao system comprising a
high order \wfs and a tip-tilt sensor.  In \darc, one thread would be
assigned to the tip-tilt sensor, which has very little work to do,
while a number of threads would be assigned to the high order \wfs,
each of which will have more computational demand.  There will also be
a post-processing thread which is used for operations that are not
suitable for multi-threaded use, for example sending mirror demands to
a deformable mirror.  In this case, lowest latency will be achieved by
giving highest priority to the high order \wfs processing threads.
The tip-tilt thread should be given a lower priority, and its work
will be completed during computation gaps.  The post-processing thread
should be given a higher priority, so that it will complete as soon as
all data are available, reducing latency.

The location of threads can also be specified, restricting a given
thread to run on one, or a subset of \cpu cores.  This prevents the
kernel from migrating threads to different cores, and also allows
threads to be placed closest to hardware in non-uniform systems (for
example where one \cpu has direct access to an interface bus), so
improving performance.  A fine tuning of latency and jitter can be
achieved in this way.

\subsubsection{Sub-aperture numbers and ordering}
Another optimisation that can be made, but which is far less obvious,
is to ensure that there are an even number of sub-apertures defined
for each wavefront sensor, and that pairs of sub-apertures are
processed by the same processing thread (ensured using the
sub-aperture allocation facility).  This ensures that wavefront slopes
are aligned on a 16 byte memory boundary for the partial matrix-vector
multiplication during the wavefront reconstruction processing, and
allows \sse operations (vector operations) to be carried out.  It
should be noted that if the system contains an odd number of
sub-apertures, an additional one can be added that has no impact on
the final \dm calculations simply by adding a column of zeros to a
control matrix.  This optimisation can have a large impact on
performance.  Where possible, all matrix and vector operations in
\darc are carried out using data aligned to a 16 byte boundary to make
maximum use of \sse operations.

\subsubsection{Pixel readout and handling}
A low latency \ao system will usually use a \wfs camera for which
pixels can be made available for processing as they are read out of
the camera, rather than on a frame-by-frame basis, or at least made
available in chunks smaller than a frame.  This allows processing to
begin before the full frame is available, and since camera readout is
generally slow, this can give a significant latency reduction, and can
mean that minimal operations are required once the last pixel
arrives.  With \darc, optimisations can be made by optimising the
``chunk'' size, i.e. the number of pixels that are made available for
processing together.  A smaller chunk size will require a larger
number of interrupts to be raised, and also a larger number of data
requests (typically \dma).  Conversely, a larger chunk size will mean
that there is a greater delay between pixels leaving the camera and
becoming available for processing.  There is therefore a trade-off to
be made, that will depend on camera type, data acquisition type and
processor performance among other things.

Related to this is an optimisation that allows \darc to process
sub-apertures in groups.  There is a parameter that is used by \darc
to specify the number of pixels that must have arrived before a given
sub-aperture can be processed.  If this value is rounded up to the
nearest multiple of chunk size then there will be multiple
sub-apertures waiting for the same number of pixels to have arrived,
and hence these can be processed together, reducing the number of
function calls required, and hence the latency.  The processing of
particular sub-apertures can be assigned to particular threads to help
facilitate this optimisation.

\subsubsection{Linux kernel impact}
The Linux kernel version with which \darc is run can also have an
impact on latency.  We do not have a definitive answer to which is the
best kernel to use, because this depends somewhat on the system
hardware, and also \ao system order.  We also find that when using a
real-time kernel (with the RT-preempt patch), latency (as well as
jitter) is slightly reduced.  Therefore, if \darc is struggling to
reach the desired latency for a given system, investigating different
kernels may prove fruitful.  This can also impact diagnostic bandwidth too.

\subsubsection{Grid utilisation}
More ambitious latency reduction can be achieved by spreading the \darc
computational load across a grid computing cluster.  For some \ao
systems, the division of work will fall naturally.  For example, systems
with more than one \wfs could place each \wfs on a separate grid
node, before combining the results in a further node which sends
commands to a \dm.  For other \ao systems, the division of labour may
not be so obvious, for example a \xao system with only a single
wavefront sensor.  In cases such as this, separation would be on a per
sub-aperture basis with responsibility for processing different
sub-apertures being placed on different grid nodes.  However, to
achieve optimal performance in this case it must be possible to split
\wfs camera pixels between nodes using for example a cable splitter,
rather than reading the \wfs into one node and then distributing
pixels from there, which would introduce additional latency.
 
The effectiveness of using \darc in a grid computing environment
depends to some extent on the communication link between grid nodes.
Real-time data must be passed between these nodes, and so we recommend
that dedicated links be used, that will not be used for other
communications, such as diagnostic data.  These links should also be
deterministic to reduce system jitter.  Additionally, the higher the
performance of these links, the lower the overall latency will be.  A
\darc implementer will need to implement their own \darc modules
according to the communication protocol and hardware that they use, as
the standard \darc package contains only a TCP/IP implementation that
is not ideal for low jitter requirements.  It will also be necessary
for the implementer to ensure that a lower latency is achieved when
using a grid of computers than can be achieved on a single computer.

\section{DARC implementations}
Optimising \darc for a particular \ao system requires some thought.
Using hardware available at Durham, consisting of a dual six-core
processor server (Intel E5645 processors with a clock speed of
2.4~GHz) with three NVIDIA Tesla 2070 \gpu acceleration cards, we have
implemented the real-time control component of several cutting edge
existing and proposed \ao systems, as given in
Table~\ref{tab:implementations}.  In doing this, we have sought to
minimise the latency that can be achieved using \darc for these
systems.  In the following sections we discuss these implementations,
the achievable performance and the implications that this has.

\begin{table}
\tiny{
\begin{tabular}{lllll}\hline
System & Wavefront & Deformable  & Frame & Telescope\\ 
& sensor & mirror & rate & size \\ \hline
VLT planet-finder & $40\times40$ & $41\times41$ & 2~kHz & 8~m\\
Palm-3000 & $62\times62$ & 3300 & 2~kHz & 5~m\\
ELT-EAGLE & 11 of $84\times84$ & 20 of $85\times85$ & 250~Hz & $\sim$40~m\\ \hline
\end{tabular}
}
\caption{A table detailing some existing and proposed AO systems with
  demanding computational requirements for the real-time system.  The
  high degree of correction make these systems cutting-edge in the
  science they can deliver.}
\label{tab:implementations}
\end{table}

\subsection{DARC as a RTC for system resembling a VLT planet-finder}
A planet-finder class instrument is currently under development for
one of the \vlt telescopes in Chile.  The \ao system for this
instrument, SPHERE \citep{sphere}, is based on a $40\times40$ sub-aperture
Shack-Hartmann extreme \ao system.  Real-time control will be provided
by an \eso standard \sparta system, comprised of a Xilinx Virtex
II-Pro \fpga for pixel processing (\wfs calibration and slope
calculation), and a DSP based system for wavefront reconstruction and
mirror control, comprised of four modules of eight Bittware Tigershark
TS201 DSPs.

This \ao system is required to operate at a frame rate of at least
1.2~kHz, and a goal of 2~kHz, with a total latency (including detector
readout and mirror drive) of less than 1~ms \citep{sphere}.

We have implemented an equivalent \ao \rtcs using \darc, based on the
aforementioned server PC, but without using the \gpu acceleration
cards.  It should be noted that we do not have an appropriate camera
or \dm to model this system, and so the implementation here does not
include these system aspects, however it does include all other
aspects of a real-time control system, including interleaved
processing and readout and thread synchronisation.  We are able to
operate this system at a maximum frame rate of over 3~kHz, using 12
CPU threads, corresponding to a total mean frame processing time of
about 323~$\mu$s, as shown in Fig.~\ref{fig:nthreads}.  In a real
system (with a real camera), \wfs readout would be interleaved with
processing, and so the latency, defined in this paper as the time
taken from last pixel received to last \dm actuator set, would be
significantly less than this, because most processing would occur
while waiting for pixels to be read out of a camera and sent to the
\rtcs.  Therefore, the latency measured from last pixel acquired to
last command out of the box is expected to be well below 100~$\mu$s,
well within the required specification.  Here, we find that processing
sub-apertures in blocks of about 40 at a time gives best performance.
When interpreting these results, it should be noted that in a standard
configuration as used here, \darc does not allow pipelining of frames:
computation of one frame must complete before the next frame begins,
and so the maximum frame rate achievable is the inverse of the frame
computation time.

\begin{figure}
\includegraphics[width=8cm]{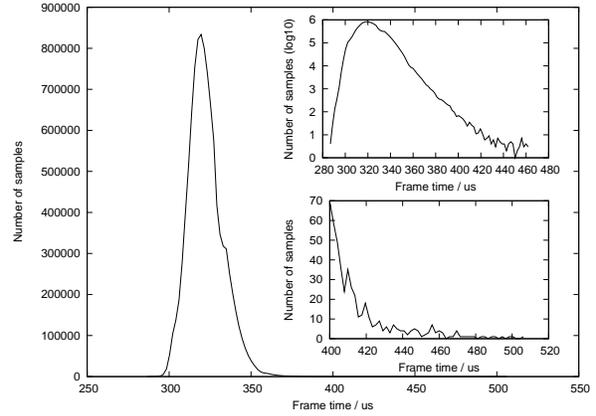}
\caption{A histogram of frame computation times for a 40x40
  sub-aperture AO system measured with $10^7$ samples.  Inset is shown
  a logarithmic histogram (showing outliers more clearly), and also a 
  linear histogram of outliers only.  This clearly shows that jitter
  is well constrained.}
\label{fig:hist}
\end{figure}

\darc performance for this configuration was assessed by using the Linux
real-time clock to measure frame computation time.  A real-time Linux
kernel (2.6.31-10.rt, available from Ubuntu archives) was used.  The
frame computation time was measured to be $323\pm11$~$\mu$s,
averaged over ten million consecutive frames, a histogram of which is
shown in Fig.~\ref{fig:hist}.  The system jitter of
11~$\mu$s \rms was measured (the standard deviation of frame
computation time).  The maximum frame time measured over this period
was 508~$\mu$s, though this (as can be seen from the standard deviation)
was an extremely rare event.  In fact, only 501 frames (out of $10^7$)
took longer than 400~$\mu$s to compute, only 106 frames took longer
than 425~$\mu$s, and only 2 frames took longer than 500~$\mu$s.  The
mean frame computation time here corresponds to a frame rate of
greater than 3~kHz.  

Equivalent measurements made with a stock Linux kernel
(2.6.32) did not give significantly worse performance, with the mean
frame computation time increasing slightly to $343\pm13$~$\mu$s and no
increase in the processing time tail.  

The introduction of a camera and a \dm to this system would increase
the frame computation time due to the necessity to transfer pixel data
and \dm demand data, however our experience shows that we would not
expect a significant increase in computation time, and maximum frame
rates greater than 2~kHz would still be easily achievable.

It is interesting to note that more recent stock Ubuntu kernels
(2.6.35 and 2.6.38) give far worse performance, almost doubling the
frame computation time.  At this point we have not investigated
further, though intend to do so.  This could be due to parameters used
during kernel compilation, or due to actual changes in the kernel
source, though other available bench marks do not suggest that this is
a likely problem.  However, it is worth bearing mind that performance of a
software based \ao control system may be dependent on the operating
system kernel used.

Typically, a design for an \ao real-time control system is made before
actual hardware and software is available, and so predictability of
performance of a \cpu-based real-time controller is not usually well
defined.

However, by using preexisting real-time control software such as
\darc, performance predictability can be improved, as it allows
hardware to be purchased earlier in the development cycle of a
real-time control system, for immediate use.  This removes much of the
uncertainty of \cpu-based controllers much earlier in the design and
prototyping phases of \ao system development.

\subsection{DARC as a RTC for a system resembling the Palm-3000 AO system}
The Palm-3000 \ao system on the five meter Hale telescope is the
highest order \ao system yet commissioned \citep{2008SPIE.7015E..95T},
and has the highest computational demands.
At highest order correction, it uses a $62\times62$ Shack-Hartmann
wavefront sensor and a \dm with about 3300 active elements, and is
specified to have a maximum frame rate of 2~kHz.  For real-time
control, this system uses 8 PCs and 16 \gpus,
far more hardware than we have available at our disposal.  However, by
using \darc on our existing (previously mentioned) hardware, we have
been able to meet this specification, though again, as we do not have
suitable cameras and \dms, our measurements do not include these.

In order to implement this system, we use the three Tesla \gpus for
wavefront reconstruction, spreading the matrix-vector multiplication
equally between then.  The matrix in each \gpu has dimensions equal to
$N_{\rm act} \times N_{\rm slopes}/3$, with $N_{\rm act}$ being the
number of \dm actuators, and $N_{\rm slopes}$ being the number of
slopes (twice the number of active sub-apertures).  In order to
interleave camera read-out with pixel processing, we split these
multiplications into four blocks, i.e.\ perform a partial
matrix-vector multiplication once a quarter, a half, three quarters and
all the slope measurements for each \gpu have been computed, thus
performing a total of twelve matrix-vector multiplications per frame.
It is interesting to note that the actual Palm 3000 \rtcs splits
processing into two blocks per \gpu, i.e.\ processing occurs half way
through pixel arrival and after all pixels have arrived.  Our
implementation performs pixel calibration and slope calculation
in \cpu, dividing the work equally between twelve processor cores
using twelve processing threads.  

By configuring \darc in this way, we are able to achieve a frame rate
of 2~kHz and a corresponding frame processing time of 500~$\mu$s.  The
addition of a real camera and \dm would increase this processing time
meaning that the official specifications would not be met.
However by adding a fourth \gpu, performance could be increased.

It should be noted that our \gpus are of a higher specification, with
memory bandwidth (the bottleneck) being 65~\% higher than those used
by Palm-3000, which will account for some of the difference.  We also
perform calibration and slope computation in \cpu, while the Palm-3000
system performs this in \gpu, and the Palm-3000 system will include
some overhead for contingency (i.e.\ the maximum frame-rate is likely
to be greater than 2~kHz should the wavefront sensor allow it).
Because all our calculations are performed within one computing node,
we do not suffer from increased latency due to computer-computer
communication, which the Palm-3000 system will include.  These
differences help to explain how \darc is able to perform the same task
using significantly less hardware and a simpler design.

This demonstrates that \darc is suitable for use with high order \ao
correction, despite being primarily \cpu based, i.e.\ \cpu based
real-time control systems are sufficiently powerful for current \ao
systems.  

\subsection{DARC as a RTC for a system resembling E-ELT EAGLE}
\label{sect:eagle}
\begin{figure}
\includegraphics[width=8cm]{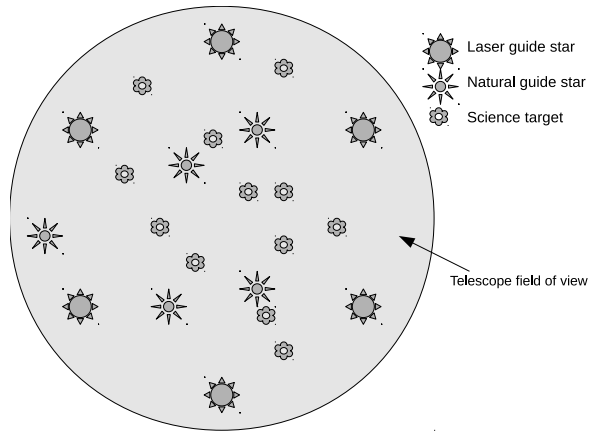}
\caption{A figure demonstrating multi-object adaptive optics.  The
  large circle represents the telescope field of view, and six laser
  guide stars and associated wavefront sensors are arranged in a ring
  near the edge of this.  Five natural guide star wavefront sensors
  are then placed on appropriate stars, and multiple science targets,
  each with a deformable mirror can then be picked out.}
\label{fig:moao}
\end{figure}

A multi-object spectrograph for the \eelt is currently in the design
phase.  This instrument, \eagle \citep{2008SPIE.7014E..53Cshort}, will have
a multi-object \ao system \citep{2010SPIE.7736E..25Rshort}, with independent wavefront correction along
multiple lines of sight, in directions not necessarily aligned with
wavefront sensors, as shown in Fig.~\ref{fig:moao}.  A current design
for \eagle consists of 11 \wfss (of which six use laser guide stars),
and up to 20 correction arms.  Each \wfs has $84\times84$
sub-apertures, and each correction arm contains an $85\times85$
actuator \dm, updated at a desired rate of 250~Hz.

The real-time control requirement for this system is demanding, though
each correction arm is decoupled, and hence can treated as a separate
\ao system.  Therefore each of these systems will have 11 $84\times84$
\wfss, and one $85\times85$ actuator \dm, leading to a total of about
125000 slope measurements and 5800 active actuators.  A control matrix
for this system would have a size of nearly 3~GB (assuming elements
are stored as 32 bit floating point).  Generally, for a system of this
size, processing power is not the limiting factor.  Rather, it is the
memory bandwidth required to load this control matrix from memory into
the processing units for each frame (be they CPUs, GPUs, FPGAs etc.),
which in this case is equal to about 725~GB/s (matrix size multiplied
by frame rate).  The 3-GPU system that we have in Durham has a peak
theoretical bandwidth of 444GB/s, and our matrix-vector multiplication
core is able to reach about 70\% of this.  Theoretical, or even
measured matrix multiplication rates are however not a good bench mark
for a real-time control system.  This is because the \rtcs will
perform additional operations, which will affect cache or resource
usage, and furthermore, the multiplication will be broken up into
blocks allowing reconstruction to be interspersed with pixel readout.
\darc is an ideal tool for such benchmarking, as it is both a full
\rtcs, but also flexible enough to investigate parameters for optimal
performance.

\begin{figure}
\includegraphics[width=8cm]{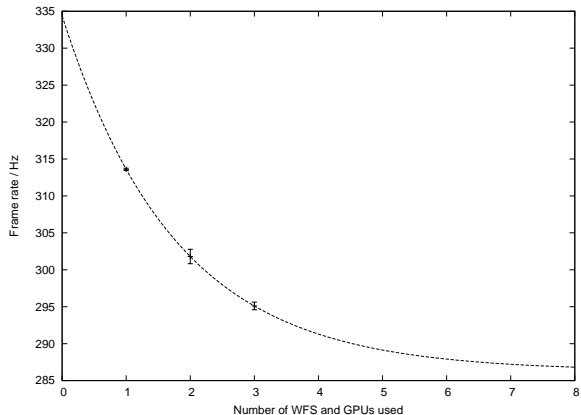}
\caption{A figure showing how frame rate is affected by the number of
  wavefront sensors processed for an EAGLE-like system, assuming a GPU
  dedicated to each wavefront sensor all on the same computer.  An
  exponential trend line has been fitted to the data.}
\label{fig:eagleframerate}
\end{figure}

By using the three Tesla \gpus that we have available, we are able to
process one wavefront sensor on each \gpu at a frame rate of about
300~Hz, with wavefront reconstruction for one correction arm
(i.e. $\frac{3}{11}^{\rm th}$ of a single channel).  We find that the
frame rate falls slightly with the number of \wfss processed (and
hence the number of \gpus used) as shown in
Fig.~\ref{fig:eagleframerate}.  If we consider how a system
containing eight \gpus might behave (the maximum number of \gpus that
can be placed in a single PC), then an extrapolation to eight \wfss and
\gpus (which we realise is rather dubious from the available data, but
will suffice for the argument being made here) might bring the frame
rate down to about 280~Hz.  For a single channel of \eagle,
reconstruction from eleven \wfss is required, which if divided between
eight \gpus would reduce the frame rate to about 200Hz.  If we were to
change the \gpus used from Tesla 2070 cards to more powerful (yet
cheaper) GeForce 580 cards (increasing the \gpu internal memory
bandwidth from 148~GBs$^{-1}$ to 192~GBs$^{-1}$), the frame-rate could
be increased to 260~Hz.

Since \eagle channels are essentially independent, we could then
replicate this system using a single computer and eight \gpus for each
channel (and one more to control the \eelt deformable M4 mirror with
$85\times85$ actuators), giving a full real-time solution for \eagle.
In fact, this situation is even simpler, since we only need to perform
wavefront slope measurement once per frame, rather than once per
channel per frame, and so a front end (possibly \fpga based for
minimal latency, for example the \sparta wavefront processing unit)
could be used to compute wavefront slopes, which would greatly reduce
the \cpu processing power required for each channel (though \gpu
processing requirement would remain unchanged).

It should be noted that in the case of wavefront reconstruction only
(assuming slope calculation is carried out elsewhere), then our system
in Durham is able to achieve a frame rate of 400~Hz, independent of
whether we process slopes from one, two or three \wfss using a \gpu
for each.  We are therefore confident that such a system would allow
us to implement the entire \eagle real-time control system using
currently available hardware and software.  Given the performance
improvements promised in both multi-core \cpus and in \gpus over the
next few years, a \cpu and \gpu based solution for \eagle is even more
feasible.

\subsection{Future real-time control requirements}
In addition to the consideration of real-time control for \eagle
given in \S\ref{sect:eagle}, we should also consider other future
real-time requirements, whether \darc will be able to meet these, and
what the \ao community will require.  

The most demanding currently proposed instrument (in terms of
computational requirement) is probably \epics
\citep{2007lyot.confE..35K} for the \eelt.  This \xao system consists
of a \wfs with $200\times200$ sub-apertures and a frame rate of 2~kHz.
Using \darc with currently available \gpus (NVIDIA GeForce 580 with a
memory bandwidth of 192~GBs$^{-1}$) performing matrix-vector
multiplication based wavefront reconstruction would require a system
with at least 150 \gpus.  Although this is a similar number to that
required by \eagle, for \epics the results from each \gpu must be
combined with results from all other \gpus, and to the authors, this
does not seem to be a practical solution when a frame rate of 2~kHz is
required (for \eagle, only computations from sets of eight \gpus need
to be combined since the \moao arms can be treated independently).  On
this scale, other reconstruction algorithms do not seem appropriate:
Conjugate gradient algorithms cannot be massively parallelised to this
degree, and, as far as we are aware, algorithms such as CuRe
\citep{cure} do not yet provide the correction accuracy required.
Therefore, we must conclude that \darc is not suitable for this
application.  However, given that this instrument is at least 10 years
away, more powerful hardware and more suitable algorithms are likely
to become available.

To our knowledge, other proposed instruments generally have lower
computational and memory bandwidth requirements than \eagle, which we
have shown that \darc would be capable of controlling.  Therefore we
are confident that \darc provides a real-time control solution for
most proposed \ao systems, and this demonstrates the suitability of
\cpu based real-time control systems for \ao.  It should be noted that
it is only within the last few years, with the advent of multi-core
\cpus and more recently \gpu acceleration that such systems have
become feasible, giving the advantages of both greater processing
power, and additional \cpus to handle non-real-time processes (such as
operating system services and diagnostic systems), thus keeping jitter
to a minimum.

\subsection{Wavefront sensor camera and deformable mirror specifications}
The timing measurements for \darc provided here are optimistic since
we do not include a physical wavefront sensor camera or \dm.  However,
by considering the latency and frame-rate requirements for the systems
for which we have investigated the performance of \darc, we can derive
the specifications required for these hardware components that will
allow the system to perform as desired.

A frame computation time of 323~$\mu$s was measured for \darc
operating in a VLT planet-finder configuration.  The total latency for
this system (including camera readout) must be below 1~ms.  However,
when camera readout time, and \dm settling time is taken into account,
this corresponds to an acceptable real-time control system latency
(from last pixel received to last \dm command leaving) of about
100~$\mu$s (E. Fedrigo, private communication).  To achieve this
latency, we can therefore specify that the camera pixel stream must be
accessible in blocks that are equal to or less than quarter of an
image in size (the time to process each block will then be about
80~$\mu$s, meaning that it will take this long to finish computation
once the last block arrives at the computer, resulting in a 80~$\mu$s
\rtcs latency.

The \darc configuration for a system resembling Palm-3000 provides a
processing time of 500~$\mu$s using three \gpus.  Once a real \wfs
camera and \dm are added, this processing time would increase meaning
that the 2~kHz frame rate could no longer be met.  Therefore, a fourth
\gpu is required to be added to the system, which would result in a
processing time of less than 400~$\mu$s.  This therefore provides a
100~$\mu$s overhead to allow for the process of sending commands to
the \dm and obtaining pixel data from a frame grabber card, which
should be more than ample (typically this will just be a command to
initiate a \dma), thus putting a performance requirement on the
hardware.  The maximum camera frame rate is 2019~Hz, corresponding to
a readout time of 495~$\mu$s.  This is greater than the processing
time, meaning that we are therefore able to interleave processing with
readout, with enough time to finish block processing between each
block of pixels arriving.  The \rtcs latency (last pixel received to
last \dm command sent) will therefore be determined by the pixel block
size.  If we read the camera in four blocks (giving a 100~$\mu$s
processing time per block) then after the last pixel arrives, we will
have a 100~$\mu$s processing time plus the 100~$\mu$s overhead that
we have allowed for data transfer, a total of 200~$\mu$s \rtcs
latency.  For this system, we therefore require that the camera pixels
can be accessed in four blocks, and that the time overhead for transferring
pixels into memory and commanding the \dm is less than 100~$\mu$s.

Our configuration of \darc for an \eagle-like instrument gives a
processing time of 3.8~ms. To achieve the desired frame rate of
250~Hz, the total processing time must remain below 4~ms.  We
therefore have 200~$\mu$s in which to process the last block of pixels
(processing of other blocks will be interleaved with pixel readout and
thus will not contribute to the \rtcs latency), and to send commands
to the deformable mirror.  If we divide the pixel data for each camera
into 42 blocks (which corresponds to two rows of sub-apertures per
block), then the processing time for each block is 90~$\mu$s, thus
leaving 110~$\mu$s spare.  We have therefore placed requirements on
the \dm and \wfs camera for this system: We require a camera which
will allow us to access the pixel stream in blocks of size 1/42 of a
frame.  We also require the time to receive this block of pixel data
into computer memory, and to send \dm demands from computer memory to
be less than 100~$\mu$s.  This equates to a data bandwidth of about
4~GBits~s$^{-1}$, which is achievable with a single PCI-Express
(generation 2) lane.  This requirement can easily be met given that
camera interface cards typically use multiple lanes, and a \dm
interface card could also use multiple lanes.

\section{Conclusion}
We have presented details of the Durham \ao real-time control system,
including recent improvements.  We have described some of the more
advanced features and algorithms available with \darc and given
details about the improved modularity and flexibility of the system
(including the ability to load and unload modules while in
operation).  We have also discussed ways in which \darc can be
optimised for specific \ao systems.  

We have discussed how \darc can be used for almost all currently
proposed future \ao systems, and given performance estimates for some
of these.  We are well aware that without a physical camera and \dm,
the results that we have presented here do not represent the whole
system, and so we have used these performance estimates to derive the
wavefront sensor camera and \dm requirements that are required to
ensure an \ao system can meet its design performance.  The
architecture of \darc, allowing interleaving of processing and camera
readout means that \ao system latency can be kept low, and so we are
confident that \darc presents a real-time control solution that is
well suited to \elt scale systems.

By making use of \gpu technology, we have been able to provide a
real-time control system suited for all but the most ambitious of
proposed \elt instruments, and have demonstrated that software and
\cpu based real-time control systems have now come of age.

\section*{Acknowledgements}
This work is funded by the UK Science and Technology Facilities
Council.  The authors would like to thank the referees who helped to
improve this paper.

\bibliographystyle{mn2e}

\bibliography{mybib}
\bsp

\end{document}